\documentclass[a4paper,11pt]{article}
\usepackage{pos}
\usepackage{amsmath}
\newcommand{\angstrom}{\textup{\AA}}

\graphicspath{{./plots/}}

\title{High-precision Quantum Monte-Carlo study of charge transport in a lattice model of molecular organic semiconductors}
\ShortTitle{Lattice model of organic semiconductors}

\author*[a]{Pavel Buividovich}
\author[b]{Johann Ostmeyer}
\author[c]{Alessandro Troisi}

\affiliation[a]{Department of Mathematical Sciences, University of Liverpool,\\
  Liverpool, L69 7ZL, UK}

\affiliation[b]{Helmholtz-Institut f\"{u}r Strahlen- und Kernphysik, Rheinische Friedrich-Wilhelms-Universit\"{a}t Bonn,\\ 53115 Bonn, Germany}

\affiliation[c]{Department of Chemistry, University of Liverpool,\\
		Liverpool, L69 7ZD, UK}

\emailAdd{pavel.buividovich@liverpool.ac.uk}

\abstract{We use first-principle Quantum Monte-Carlo (QMC) simulations and numerical exact diagonalization to analyze the low-frequency charge carrier mobility within a simple tight-binding model of molecular organic semiconductors on a two-dimensional triangular lattice. These compounds feature transient localization, an unusual charge transport mechanism driven by dynamical disorder. The challenges of studying the transient localization of charge carriers in the low-frequency/long-time limit from first principles are discussed. We demonstrate that a combination of high-precision QMC data with prior estimates of frequency-dependent charge carrier mobility based on the static disorder approximation for phonon fields allows for improved estimates of mobility in the low-frequency limit. We also point out that a simple relaxation time approximation for charge mobility in organic semiconductors is not consistent with the QMC data. Physical similarities with charge transport in quark-gluon plasma are highlighted.}

\FullConference{The 41st International Symposium on Lattice Field Theory (LATTICE2024)\\
 28 July - 3 August 2024\\
Liverpool, UK\\}

\newcommand{\lr}[1]{ \left( #1 \right) }
\newcommand{\lrs}[1]{ \left[ #1 \right] }
\newcommand{\lrc}[1]{ \left\{ #1 \right\} }
\newcommand{\abs}[1]{ \left| #1 \right| }
\newcommand{\vev}[1]{ \left\langle \, #1 \, \right\rangle }
\newcommand{\cev}[1]{ \left\langle \left\langle \, #1 \, \right\rangle \right\rangle }

\newcommand{\ket}[1]{ \, | #1 \rangle }
\newcommand{\bra}[1]{ \langle #1 | \, }

\newcommand{\tr}{ {\rm Tr} \, }

\newcommand{\meV}{ \, {\rm meV} }

\newcommand{\expa}[1]{ \exp{\left( #1 \right)} }

\newcommand{\HH}{\hat{H}}
\newcommand{\ZZ}{\mathcal{Z}}
\newcommand{\hc}{\hat{c}}

\begin{document}
\maketitle

\section{Introduction}
\label{sec:intro}

Organic semiconductors are crystals made of organic molecules that are held together by inter-molecular Van der Waals forces. While rubrene and pentacene are some of the best known examples, this class of materials includes thousands of compounds, many of which have never been synthesized. Organic semiconductors are promising candidates for the development of organic ink-jet-printable electronics (e.g. displays, solar cells, flexible devices, biosensors) with lower cost and larger area than silicon-based devices.

Despite a great variety of organic semiconductors, most of them have layered structure, where charge transport mostly happens within two-dimensional layers and hoppings between layers are negligible. In most cases, charge transport within each layer can be described in terms of a simple tight-binding Hamiltonian describing electron-phonon or hole-phonon interactions on a two-dimensional triangular lattice \cite{TroisiNatureMaterials2017}:
\begin{eqnarray}
	\label{eq:full_Hamiltonian}
	\hat{H}_{F} 
	= 
	\sum\limits_{x,y} \hc^{\dag}_{x,s} h_{xy}\lrs{\hat{\phi}_{z,A}} \hc_{y,s}
	+
	\hat{H}_B, 
	\quad
	\hat{H}_B = \sum\limits_{z,A} \lr{\frac{\hat{\pi}_{z,A}^2}{2} + \frac{\omega_0^2 \, \hat{\phi}_{z,A}^2}{2}} ,
	\\
	\label{eq:single_particle_Hamiltonian}
	h_{xy}\lrs{\hat{\phi}_{z,A}} 
	= 
	-\kappa_A \lr{1 - \lambda_A \hat{\phi}_{x,A}} \delta_{x+A,y} 
	-
	\kappa_A \lr{1 - \lambda_A \hat{\phi}_{x-A,A}} \delta_{x-A,y} .
\end{eqnarray}
Here $\hc^{\dag}_{x,s}$, $\hc_{x,s}$ are the creation/annihilation operators for fermionic charge carriers with spin $s = \uparrow, \downarrow$. Lattice sites are labelled by indices $x$, $y$, $z$, and $A = 1 \ldots 3$ labels the three (forward) lattice bonds (links) attached to each site. $h_{xy}\lrs{\hat{\phi}_{z,A}}$ is a single-particle Hamiltonian for charge carriers, where $\kappa_A$ are the hopping amplitudes (transfer integrals) along lattice bonds in the direction $A$, and $x \pm A$ denotes lattice sites connected to $x$ by a bond in forward or backward direction $A$. The hopping amplitudes are modulated by phonon modes $\hat{\phi}_{z,A}$ associated with each bond ($\hat{\pi}_{z,A}$ is the corresponding canonically conjugate momentum). The electron-phonon couplings $\lambda_A$ control the strength of this modulation. With a very good precision, the phonon Hamiltonian $\hat{H}_B$ is a sum of independent harmonic oscillators with a single frequency $\omega_0$ on all bonds. 

To discuss an unusual mechanism of charge transport in materials described by the Hamiltonian (\ref{eq:full_Hamiltonian}), let us first note that in the absence of phonons ($\phi_{x,A} = 0$) the dispersion relation for charge carriers in the conductance band is a single function $\epsilon\lr{\vec{k}}$, where $\vec{k}$ is the lattice momentum. Hence, there are no inter-band processes that could contribute to the optical conductivity $\sigma\lr{\omega}$ at $\omega \neq 0$, and $\sigma\lr{\omega}$ vanishes for all $\omega \neq 0$ (in contrast to e.g. Dirac materials, where there are two distinct energies $\pm \epsilon\lr{\vec{k}}$ for each $\vec{k}$ and $\sigma\lr{\omega}$ is saturated by transitions between upper and lower Dirac cones \cite{Buividovich:12:1}). Hence non-vanishing optical conductivity is only possible in the presence of electron-phonon interactions. Since inter-molecular Van der Waals forces in organic crystals are usually rather weak, typical phonon frequencies $\omega_0$ in molecular organic semiconductors are around $5 \meV$, considerably smaller than the room temperature $T = 300 \, \mathrm{K} = 25 \meV$. As a result, phonon modes have large population numbers, and fluctuations of phonon-modulated hopping amplitudes $\kappa_A \lr{1 - \lambda_A \hat{\phi}_{x,A}}$ in the single-particle Hamiltonian (\ref{eq:single_particle_Hamiltonian}) are comparable to $\kappa_A$ themselves. Charge carriers hence propagate in a nearly-classical, strongly disordered phonon background. This regime is quite similar to early stages of heavy ion collisions, where quarks propagate in the background of nearly-classical soft gluon modes \cite{Gelis:1002.0333}. It is not surprising that charge transport in both organic semiconductors and out-of-equilibrium quark-gluon plasma can be described with reasonable precision in terms of quantum fermions (quarks or electrons) interacting with classical bosons (phonons or gluons) \cite{Berges:1303.5650,TroisiOrlandiPRL2006}.

Typical values of hopping amplitudes $\kappa_A \sim 50 \ldots 100 \meV$ are also much larger than typical phonon frequencies $\omega_0 \sim 5 \meV$. Thus charge carriers are considerably ``faster'' than phonons, and, in the first approximation, see the nearly classical phonon background as a static disorder. The static approximation indeed provides a reasonable description of optical conductivity of organic semiconductors at frequencies $\omega \gtrsim \omega_0$, see Fig.~\ref{fig:spf_G}. It is analogous to ``electric QCD'' and ``magnetic QCD'' effective field theories for high-temperature Quantum Chromodynamics \cite{Braaten:hep-ph/9510408,AppelquistPisarskiPRD1981MagneticQCD}.

However, in one or two dimensions any static disorder causes the low-frequency limit $\sigma\lr{\omega \rightarrow 0}$ of optical conductivity to vanish by virtue of Anderson localization. Nevertheless, in practice charge carriers appear to be only transiently localized within the time scales $\tau_{loc} \lesssim \omega_0^{-1}$ and are able to propagate along the crystal as the phonon background slowly changes, resulting in a finite value of $\sigma\lr{\omega \rightarrow 0}$. We conclude that charge transport in organic semiconductors is driven by dynamical disorder, and would be absent for static or vanishing disorder \cite{Fratini:2312.03840}.

With real-time simulations of quantum many-body systems being an NP-hard problem, first-principle studies of transient localization are a difficult task. This phenomenon happens at long time scales $t \gtrsim \omega_0^{-1}$ which are not easily accessible by first-principle numerical methods, and where semi-classics might fail \cite{GerlachRevModPhys.63.63,Wellein:cond-mat/9703041}. In these Proceedings, we present numerical estimates of charge carrier mobility in organic semiconductors based on high-precision Quantum Monte-Carlo (QMC) simulations of the Hamiltonian (\ref{eq:full_Hamiltonian}) \cite{Buividovich:23:1}. We demonstrate that despite exponential suppression of the low-frequency signatures of transient localization in QMC results, QMC data with sub-permille statistical errors \cite{Buividovich:23:1,Buividovich:24:1} can still provide useful information about it for the realistic range of model parameters. In particular, QMC can be used to constrain and validate phenomenological models of transient localization such as the relaxation time approximation (RTA).

\section{Methods: exact diagonalization, Quantum Monte-Carlo, and relaxation time approximation}
\label{sec:methods}

Charge carrier concentration $n$ in organic semiconductors is usually quite low, $n \sim 10^{-2}$ per unit lattice cell. Correspondingly, interactions between charge carriers can be neglected, and the full many-body Hamiltonian (\ref{eq:full_Hamiltonian}) can be reduced to the polaron-type Hamiltonian describing a single charge carrier interacting with phonons:
\begin{eqnarray}
	\label{eq:polaron_Hamiltonian}
	\hat{H} 
	= 
	h\lrs{\hat{\phi}_{z,A}}  + \hat{H}_B	
\end{eqnarray}
The Hamiltonian (\ref{eq:polaron_Hamiltonian}) is a projection of the full many-body Hamiltonian (\ref{eq:full_Hamiltonian}) to the subspace of states $\lrc{\ket{\Phi}}$ with $Q=1$ charge carriers ($\hat{Q} \ket{\Phi} = \ket{\Phi}$). This subspace is a direct product of a $N$-dimensional Hilbert space of a single charge carrier on a lattice with $N$ sites\footnote{Equivalently, Hilbert space of all single-particle wave functions $\psi_x$, where $x$ labels lattice sites.} and harmonic oscillator Hilbert spaces for phonons $\phi_{x,A}$ on all lattice bonds. 

With a very good accuracy, for low carrier concentrations $n$ the optical conductivity $\sigma\lr{\omega}$ is proportional to $n$. In semiconductor devices, $n$ and $\sigma\lr{\omega}$ change by several orders of magnitude depending on the doping level. A very useful quantity which is almost independent of $n$ if $n \ll 1$ is the charge mobility $\mu\lr{\omega} = \sigma\lr{\omega}/\lr{e n}$ ($e$ is the electron charge). It characterizes the velocity $v\lr{\omega} = \mu\lr{\omega} E\lr{\omega}$ acquired by a single charge carrier in a time-dependent electric field $E\lr{\omega}$.

Within the linear response approximation, $\mu\lr{\omega}$ can be expressed in terms of the eigenstates $\ket{\Phi_k}$ and eigen-energies $E_k$ of the Hamiltonian (\ref{eq:polaron_Hamiltonian}) as follows:
\begin{eqnarray}
\label{eq:mobility_full}
 \mu\lr{\omega} 
 = 
 \frac{\pi \, \hbar \, e}{2 \ZZ \, \omega}\sum\limits_{\alpha, k,l} \abs{\bra{\Phi_k} J^{\alpha}\lrs{\hat{\phi}_{z,A}} \ket{\Phi_l} }^2
 \lr{e^{-\beta E_k} - e^{-\beta E_l}} \, \delta\lr{\omega - \lr{E_l - E_k}} ,
 \\
 \label{eq:electric_current}
 \hat{J}^{\alpha}_{xy} \equiv J^{\alpha}_{xy}\lrs{\hat{\phi}_{z,A}} 
 =
 \frac{i}{\hbar} \sum\limits_A   
 d^{\alpha}_A \kappa_A 
 \lr{
 	\lr{1 - \lambda_A \hat{\phi}_{x,A}} \delta_{x+A,y} 
 	-
 	\lr{1 - \lambda_A \hat{\phi}_{x-A,A}} \delta_{x-A,y}
 } .
\end{eqnarray}
Here $\ZZ = \tr\lr{e^{-\beta \HH}}$ is the partition function, $\beta \equiv T^{-1}$, $J^{\alpha}$ is a single-particle charge carrier velocity operator in Cartesian direction $\alpha = 1, 2$ (same as electric current up to a factor of $e$), and $d^{\alpha}_A$ is a Cartesian vector of lattice bonds in direction $A$. For simplicity, we define $\mu\lr{\omega}$ in (\ref{eq:mobility_full}) as half of the trace of the full Cartesian mobility tensor $\mu^{\alpha\beta}$, $\mu = \lr{\mu^{11} + \mu^{22}}/2$.

A universally applicable first-principle approach to calculate $\mu\lr{\omega}$ is a numerical diagonalization of the Hamiltonian (\ref{eq:polaron_Hamiltonian}), where the eigenstates $\ket{\Phi_k}$ in (\ref{eq:mobility_full}) are directly calculated using linear algebra methods \cite{Prelovsek:1111.5931}. However, typical phonon occupation numbers are large at room temperature $T \gg \omega_0$, and a large number of phonon basis states is needed to achieve convergence. This makes the high-occupancy regime very challenging for exact diagonalization and limits realistic system sizes to $N \lesssim 5$ lattice sites, in contrast to e.g. spin chains where Hamiltonians with $N \sim 30$ degrees of freedom can be diagonalized \cite{Schnack2023ExactDiagonalization}.

Another universally applicable approach is the Quantum Monte-Carlo (QMC) method, which is free of fermionic sign problem for the Hamiltonian (\ref{eq:full_Hamiltonian}). QMC simulations produce imaginary-time correlators of charge carrier velocities, which are related to $\mu\lr{\omega}$ via the Green-Kubo relations:
\begin{eqnarray}
	\label{eq:GE_def}
	G\lr{\tau} = \frac{1}{2 \ZZ} \tr\lr{\hat{J}^{\alpha} e^{-\tau \HH} \hat{J}^{\alpha} e^{-\lr{\beta - \tau} \HH}} ,
	\quad
	G\lr{\tau} = \int\limits_0^{+\infty} d\omega \, \frac{\omega \, \cosh\lr{\omega \, \lr{\tau - \frac{\beta}{2}}}}{\pi \, \hbar \,  \sinh\lr{\frac{\omega \, \beta}{2}}} \, \mu\lr{\omega} .
\end{eqnarray}
In the absence of any prior information about $\mu\lr{\omega}$, this function can only be extracted from $G\lr{\tau}$ with rather coarse frequency resolution $\Delta \omega \sim \pi \, T \approx 80 \, \meV$ (at room temperature $T = 25 \meV$) \cite{Meyer:1104.3708}. Plots of $\mu\lr{\omega}$ in Fig.~\ref{fig:spf_G} suggest that most important low-frequency features of $\mu\lr{\omega}$, such as the displaced Drude peak \cite{Fratini:2312.03840}, cannot be reliably captured with such a coarse frequency resolution.

Previous QMC studies \cite{Mishchenko:1408.5586,deCandia:1901.03223,Wang:2203.12480} mostly considered 1D version of the Hamiltonian (\ref{eq:polaron_Hamiltonian}), where localization properties are different from realistic 2D systems, and used Diagrammatic and Worldline QMC algorithms. In \cite{Buividovich:23:1} we adopted the Hybrid Monte-Carlo (HMC) approach \cite{Blankenbecler:PhysRevD.24.2278} to simulate a single charge carrier interacting with phonons as described by the 2D Hamiltonian (\ref{eq:polaron_Hamiltonian}). In contrast, previous applications of HMC to electron-phonon interactions mostly considered thermal equilibrium states of full many-body Hamiltonians similar to (\ref{eq:full_Hamiltonian}), for which the contribution of the single-particle Hilbert space with $Q = 1$ is strongly suppressed at low charge carrier concentrations. The approach of \cite{Buividovich:23:1} is based on the representation of the correlator (\ref{eq:GE_def}) in terms of a path integral over all configurations of phonon fields $\phi_{x,A}\lr{\tau}$ in imaginary (Euclidean) time $\tau \in \lrs{0, \beta}$:
\begin{eqnarray}
\label{eq:GE_path_integral}
	G\lr{\tau} = 
	\frac{1}{2 \ZZ} 
	\int \mathcal{D}\phi_{x,A}\lr{\tau} \,
	e^{-S_B\lrs{\phi_{x,A}\lr{\tau}}} \,
	\tr U\lr{0, \beta}	
	\times \nonumber \\ \times \,
	\frac{
		\tr\lr{J^{\alpha}\lrs{\phi_{x,A}\lr{0}} U\lr{0, \tau} J^{\alpha}\lrs{\phi_{x,A}\lr{\tau}} U\lr{\tau, \beta} }
	}{\tr U\lr{0, \beta}}	,
    \\
\label{eq:phonon_action}
    S_B\lrs{\phi_{x,A}\lr{\tau}} = \frac{1}{2} \sum\limits_{x,A} \int\limits_0^{\beta} d\tau \lr{\lr{\frac{d \phi_{x,A}}{d \tau}}^2 + \omega_0^2 \, \phi_{x,A}^2}
\end{eqnarray}
where $U\lr{\tau_1, \tau_2} = \mathcal{T} \expa{-\int\limits_{\tau_1}^{\tau_2} d\tau h\lrs{\phi_{x,A}\lr{\tau}} }$ is a time-ordered exponential of a single-particle Hamiltonian in the background of phonon fields $\phi_{x,A}\lr{\tau}$. Configurations of $\phi_{x,A}\lr{\tau}$ are generated with probability proportional to $e^{-S_B\lrs{\phi_{x,A}\lr{\tau}}} \tr U\lr{0, \beta}$ using HMC. 

Since $h\lrs{\phi_{x,A}\lr{\tau}}$ is linear in $\phi_{x,A}\lr{\tau}$, the time-ordered exponentials $U\lr{\tau_1, \tau_2}$ and hence both the numerator and the denominator in the second line of (\ref{eq:GE_path_integral}) behave as exponentials of $\phi_{x,A}\lr{\tau}$. We include $\tr U\lr{0, \beta}$ into the path integral weight along with the multivariate normal distribution $p\lrs{\phi_{x,A}\lr{\tau}} \sim e^{-S_B\lrs{\phi_{x,A}\lr{\tau}}}$ and normalize the velocity correlator in the second line of (\ref{eq:GE_path_integral}) by $\tr U\lr{0, \beta}$ in order to mitigate this exponential dependence. Already for a single normally distributed variable $\phi$ with $p\lr{\phi} \sim e^{-\phi^2/2}$, exponentials $e^{\alpha \phi}$ have a heavy-tailed log-normal distribution \cite{Kaplan:1106.0073}. Monte-Carlo estimates of $\vev{e^{\alpha \phi}}$ suffer from slow convergence because dominant contributions to $\vev{e^{\alpha \phi}}$ come from non-zero $\phi = \alpha$, whereas naive Monte-Carlo samples are distributed around $\phi = 0$. An analogous shift of $\phi_{x,A}\lr{\tau}$ is a polarization of medium by the charge carrier. Including the factor $\tr U\lr{0, \beta}$ into the path integral weight takes medium polarization into account already at the level of Monte-Carlo samples. At the same time, normalizing the velocity correlators in (\ref{eq:GE_path_integral}) by $\tr U\lr{0, \beta}$ removes their exponential dependence on $\phi_{x,A}\lr{\tau}$. Normalized velocity correlators become normally distributed and statistical errors are dramatically reduced.

While the path integral weight $\tr U\lr{0, \beta} e^{-S_B}$ is not positive in general, negative values only occur for phonon field configurations where $\phi_{x,A}\lr{\tau}$ has very strong $\tau$ dependence. At high temperatures $T \sim O\lr{10 \meV}$ that are relevant for practical applications, these configurations are strongly suppressed, and we never encounter negative weights in our simulations. 

Successful phenomenological models of transient localization that capture the low-frequency features of $\mu\lr{\omega}$ reasonably well can also be constructed based on the static approximation for the Hamiltonian (\ref{eq:polaron_Hamiltonian}), where phonons $\hat{\phi}_{x,A}$ are considered as time-independent random classical fields. The mobility $\mu\lr{\omega}$ can then be expressed in terms of the eigen-energies and eigenstates $\epsilon_k$, $\ket{\psi_k}$ of a single-particle Hamiltonian $h\lrs{\phi_{x,A}}$ in the background of random phonon fields $\phi_{x,A}$. However, by virtue of Anderson localization, all eigenstates $\ket{\psi_k}$ are localized in a static random background, and $\mu\lr{\omega \rightarrow 0}$ vanishes. Slow time evolution of phonon fields is taken into account by introducing a phenomenological relaxation time parameter $\tau_{loc}$ which effectively broadens transitions between different energy levels \cite{Fratini:1505.02686,Fratini:2312.03840} and yields non-zero charge mobility at $\omega \rightarrow 0$. The resulting relaxation time approximation (RTA) expression for the mobility is
\begin{eqnarray}
\label{eq:mobility_rta}
	\mu_{RTA}\lr{\omega} = \frac{\hbar \, e}{2 \, Z \, \omega \, \tau_{loc}} \tanh\lr{\frac{\beta \omega}{2}} \, 
	\cev{  
		\sum\limits_{k,l,\alpha} \abs{\bra{\psi_k} J^{\alpha} \ket{\psi_l}}^2 
		\frac{e^{-\beta \epsilon_k} + e^{-\beta \epsilon_l}}{\lr{\omega - \epsilon_k - \epsilon_l}^2 + \tau_{loc}^{-2}}
	}	,
\end{eqnarray}
where $Z = \cev{\sum_k e^{-\beta \epsilon_k}}$ and $\cev{ \ldots }$ is an average over normally distributed static phonon fields with characteristic dispersions obtained from electronic structure calculations \cite{TroisiJChemPhys2020}. In the limit $\tau_{loc}^{-1} \rightarrow 0$, equation (\ref{eq:mobility_rta}) is a static approximation to $\mu\lr{\omega}$. With $\tau_{loc}^{-1} \sim \omega_0$, $\mu_{RTA}\lr{\omega \rightarrow 0}$ yields reasonable predictions for charge mobilities in real materials \cite{TroisiNatureMaterials2017}. The eigensystem $\ket{\psi_k}$, $\epsilon_k$ and hence $\mu_{RTA}\lr{\omega}$ can be calculated in polynomial time, in contrast to the full quantum expression (\ref{eq:mobility_full}) which is NP-hard to compute. 

\section{Numerical results}
\label{sec:results}

\begin{figure*}[h!tb]
	\includegraphics[width=0.50\textwidth]{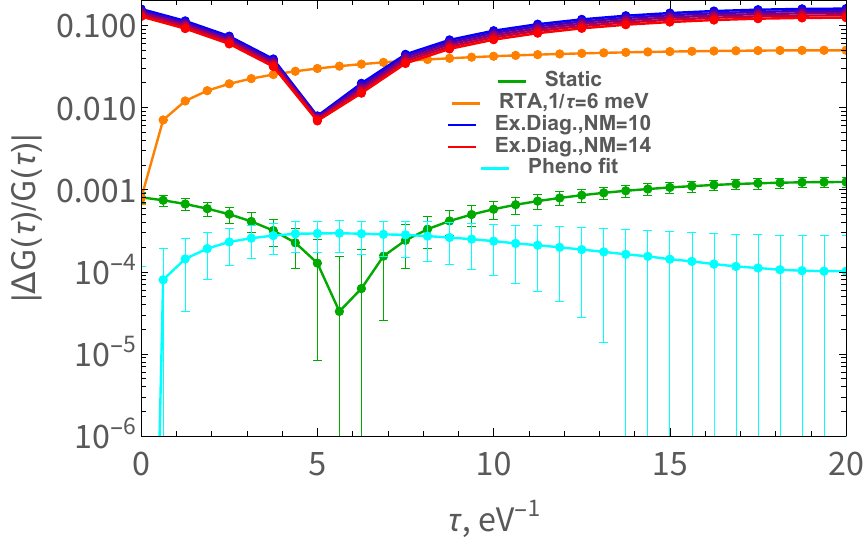}\hspace{0.02\textwidth}\includegraphics[width=0.48\textwidth]{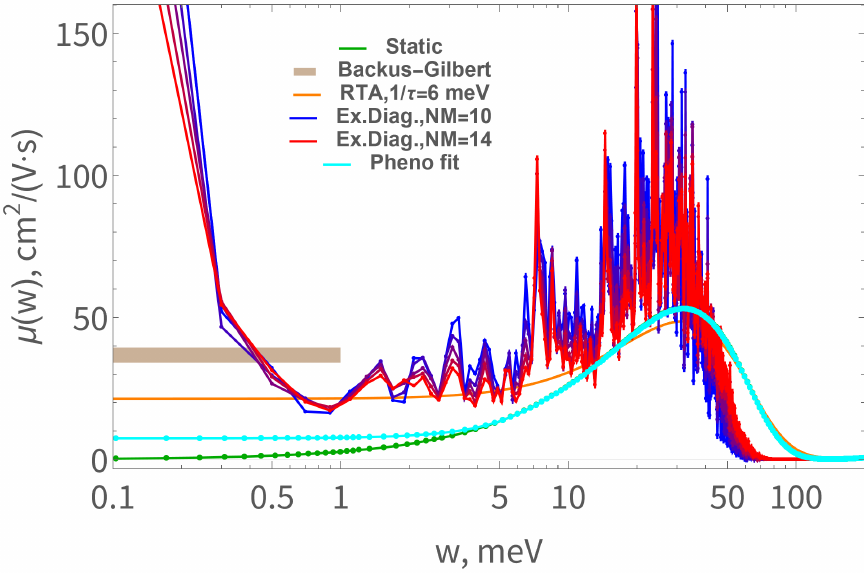}\\
	\includegraphics[width=0.505\textwidth]{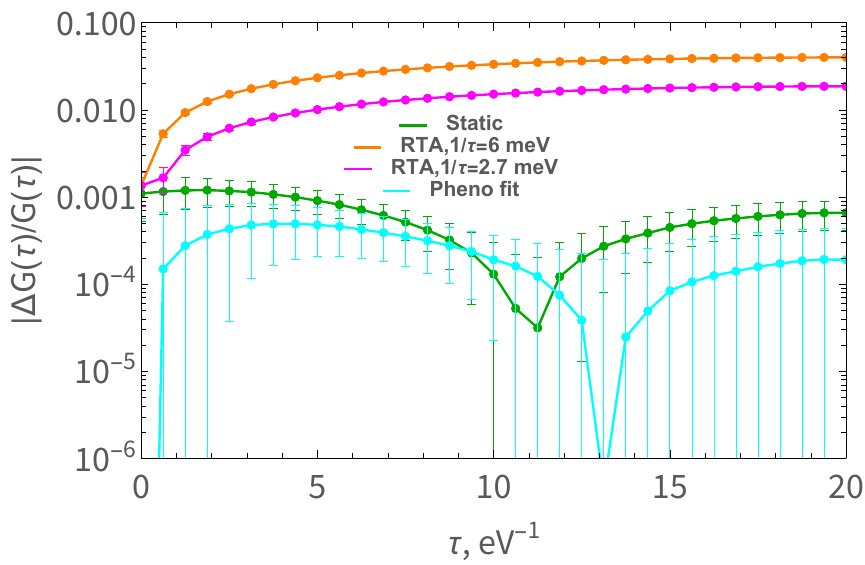}\hspace{0.02\textwidth}\includegraphics[width=0.475\textwidth]{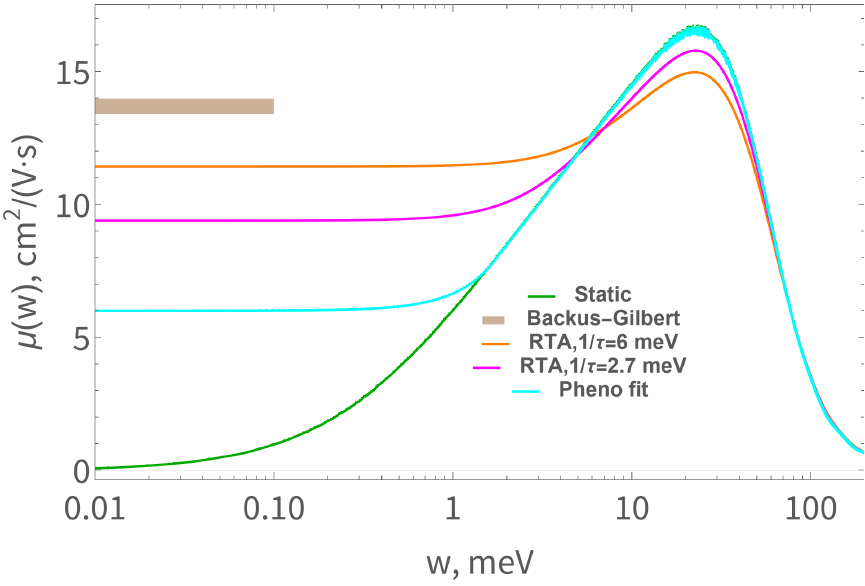}\\
	\caption{\textbf{On the left:} relative differences between full QMC results for the imaginary-time correlator (\ref{eq:GE_def}) and different approximations. \textbf{On the right:} Corresponding approximations for the frequency-dependent charge carrier mobility $\mu\lr{\omega}$ for a 1D lattice with $5$ sites (\textbf{top}) and a 2D lattice with $21 \times 21$ sites (\textbf{bottom}).}
	\label{fig:spf_G}
\end{figure*}

To compare our high-precision QMC results with exact diagonalization and static/RTA approximations, we use the Hamiltonian (\ref{eq:polaron_Hamiltonian}) with parameters that describe Rubrene, one of the best studied organic semiconductors \cite{VongJPhysChemLett2022,Buividovich:23:1}:
\begin{eqnarray}
\label{eq:rubrene_params}
 \kappa_1 = -96.1 \meV, 
 \quad 
 \kappa_2 = \kappa_3 = -14.7 \meV ,
 \quad
 \Delta \kappa_1 = 23.7 \meV, 
 \quad 
 \Delta \kappa_2 = \Delta \kappa_3 = 6.2 \meV ,
 \nonumber \\
 d_1 = \lr{7.2, 0} \angstrom,
 \quad
 d_2 = \lr{3.6, 7.15} \angstrom,
 \quad
 d_3 = \lr{3.6, -7.15} \angstrom,
 \quad
 \omega_0 = 6 \meV .
\end{eqnarray}
The corresponding values of electron-phonon couplings $\lambda_A$ are fixed by matching the dispersions of $\kappa_A \lr{1 - \lambda_A \hat{\phi}_{x,A}}$ to $\Delta \kappa_A$ in (\ref{eq:rubrene_params}). We consider a small 1D lattice with $5$ sites (with hoppings $\kappa_1$ for all bonds along $\vec{d}_1$) and a larger 2D lattice with $21 \times 21$ sites, all at room temperature $T = 25 \meV$.

We first compare QMC output with the static and RTA approximations (\ref{eq:mobility_rta}). We use the Green-Kubo relations (\ref{eq:GE_def}) to translate $\mu_{RTA}\lr{\omega}$ into the imaginary-time correlator $G_{RTA}\lr{\tau}$. Static approximation corresponds to $\tau_{loc}^{-1} \rightarrow 0$ in (\ref{eq:mobility_rta}). Plots on the left in Fig.~\ref{fig:spf_G} show that the relative difference between the full QMC result and the static approximation for $G\lr{\tau}$ is of order of $0.1 \%$. Transient localization hence manifests itself as only a tiny correction to imaginary-time QMC results, but nevertheless dominates the real-time charge carrier dynamics at low frequencies/late times. Statistical errors of our high-precision QMC results are much smaller than $0.1 \%$ and thus allow to make statistically significant predictions on the low-frequency behaviour of $\mu\lr{\omega}$.

We also observe that for both $5 \times 1$ and $21 \times 21$ lattices relative deviations of $G_{RTA}\lr{\tau}$ from the full QMC result $G\lr{\tau}$ are considerably larger for $\tau_{loc}^{-1} > 0$ than for the static approximation with $\tau_{loc}^{-1}  = 0$. In fact, any finite value of $\tau_{loc}$ increases deviations between $G_{RTA}\lr{\tau}$ and $G\lr{\tau}$. This suggests that the simple RTA approximation (\ref{eq:mobility_rta}) cannot be uniformly valid for all frequencies, and calls for more advanced phenomenology, for example, with energy-dependent $\tau_{loc}$.

Comparing the results for the $5 \times 1$ and $21 \times 21$ lattices, we see that finite-volume artefacts are very large for small lattices, however, qualitative features of $\mu\lr{\omega}$ are similar in both cases.

Next, we consider numerical exact diagonalization results for the $5 \times 1$ lattice. In the phonon Hilbert space, we use the eigenbasis of phonon occupation numbers $n_x$ truncated to $\sum_x n_x \leq N_M$. Top left plot on Fig.~\ref{fig:spf_G} demonstrates that the convergence of $G\lr{\tau}$ with respect to the occupation number cut-off $N_M$ is very slow. Even though average occupation numbers $\vev{n_x} = \lr{e^{\beta \omega_0} - 1}^{-1} = 3.69$ are considerably smaller than our values $N_M = 10 \ldots 14$, the relative difference between QMC and exact diagonalization results is as large as $10 \%$. Nevertheless, the low-frequency features of the real-frequency mobility $\mu\lr{\omega}$ (top right plot on Fig.~\ref{fig:spf_G}) appear to be rather insensitive to $N_M$. In particular, while the high-frequency part of $\mu\lr{\omega}$ approaches the static result for $\mu\lr{\omega}$ as $N_M$ increases, at intermediately small frequencies $\omega \sim 0.5 \ldots 1 \meV$ the exact diagonalization results for $\mu\lr{\omega}$ do not move down to the static approximation as $N_M$ is increased. This difference in the low-frequency behaviour of the static vs. exact diagonalization results is exactly the difference between Anderson vs. transient localization. An interesting feature is an enhancement of $\mu\lr{\omega}$ at very low frequencies $\omega \lesssim 0.5 \meV$, which has also recently been observed for 1D rubrene Hamiltonian using mapping approach to surface hopping \cite{Manolopoulos:2406.19851}. However, because of the strong $N_M$ dependence it is not clear whether this feature persists in the limit $N_M \rightarrow +\infty$. 

To provide an example of a function $\mu\lr{\omega}$ that describes the QMC data for $G\lr{\tau}$ better than the static approximation, let us construct a simple phenomenological fit with the low-frequency limit of the mobility $\mu_0 = \mu\lr{\omega \rightarrow 0}$ as a single fit parameter:
\begin{eqnarray}
	\label{eq:mu_pheno}
	\mu_F\lr{\omega} = 
	\begin{cases}
		c\lr{\mu_0} \lr{\mu_0 + \alpha\lr{\mu_0} \, \omega^2}, & \omega < \omega_1\lr{\mu_0} \\
		c\lr{\mu_0} \mu_S\lr{\omega}, & \omega \geq \omega_1\lr{\mu_0} \\ 
	\end{cases} ,
\end{eqnarray}
where $\mu_S\lr{\omega}$ is the static approximation for the mobility $\mu\lr{\omega}$. To fix the parameters $c$, $\alpha$ and $\omega_1$ given $\mu_0$, we start with $c = 1$ and find the minimal positive value of $\alpha$ such that the equation $\mu_0 + \alpha \, \omega^2 = \mu_S\lr{\omega}$ has exactly one positive solution $\omega = \omega_1$. The overall normalization $c\lr{\mu_0}$ is determined by requiring $G_F\lr{0} = G\lr{0}$. $c \neq 1$ (with $\abs{c - 1} \ll 1$) is necessary for a significant improvement of the fit in comparison with the static approximation. We then find $\mu_0$ such that
\begin{eqnarray}
\label{eq:chi2}
\chi^2 = \sum\limits_{\tau_1, \tau_2}\lr{G_F\lr{\tau_1} - G\lr{\tau_1}} C^{-1}\lr{\tau_1, \tau_2} \lr{G_F\lr{\tau_2} - G\lr{\tau_2}}
\end{eqnarray}
is minimal, where $C\lr{\tau_1, \tau_2}$ is the covariance matrix for the QMC data. Green-Kubo relations (\ref{eq:GE_def}) are used to translate $\mu_F\lr{\omega}$ into the imaginary-time correlator $G_F\lr{\tau}$. The optimal functions $\mu_F\lr{\omega}$ are shown in Fig.~\ref{fig:spf_G} (plots on the right). Plots on the left demonstrate that the corresponding deviation from $G\lr{\tau}$ becomes less than $0.05\%$ for all $\tau$ and for both 1D and 2D lattices. The optimal value of $\mu_0 = \mu\lr{\omega \rightarrow 0}$ turns out to be smaller than the RTA prediction, and also lower than the experimental value $\mu_{exp} = \lr{9.25 \pm 0.75} \mathrm{cm}^2/\lr{\mathrm{V} \cdot \mathrm{s}}$ \cite{TroisiNatureMaterials2017,VongJPhysChemLett2022}. Interestingly, the only apparent explanation for a much better agreement of $G_F\lr{\tau}$ with $G\lr{\tau}$ in comparison with $G_{RTA}\lr{\tau}$ is a small but noticeable difference in the heights of the displaced Drude peaks in $\mu_F\lr{\omega}$ and $\mu_{RTA}\lr{\omega}$ at $\omega \sim 30 \meV$. We also tried phenomenological fits of $\mu\lr{\omega}$ that include a peak at $\omega \lesssim 0.5 \meV$ similar to the one in the exact diagonalization data. However, since the relevant frequencies are very low, such fits are not restrictive enough with current statistical errors.

For comparison, we also show the estimates of $\mu\lr{\omega \rightarrow 0}$ obtained using the Backus-Gilbert method that is based solely on the QMC data \cite{Ulybyshev:1707.04212}. Because of the coarse frequency resolution $\Delta \omega \sim \pi \, T \approx 80 \meV$, Backus-Gilbert estimates of the full function $\mu\lr{\omega}$ just produce broad peaks centred at $\omega = 0$, so we do not show them to keep the plots readable. 

\section{Conclusions and discussion}
\label{sec:conclusions}

We demonstrated that a combination of high-precision QMC results from an improved HMC algorithm \cite{Buividovich:23:1,Buividovich:24:1} with prior knowledge based on the static approximation allows to extract significantly more information about the low-frequency behaviour of charge mobility $\mu\lr{\omega}$ in organic semiconductors than either QMC results or static/RTA approximations alone. It is interesting whether results obtained within electric or magnetic effective field theories of finite-temperature QCD \cite{Braaten:hep-ph/9510408,AppelquistPisarskiPRD1981MagneticQCD} can improve the reconstruction of QCD spectral functions in a similar way.

While extracting the full frequency-dependent mobility $\mu\lr{\omega}$ from QMC results is still an NP-hard problem, our QMC simulations become precise enough to test the validity of different phenomenological models and approximations for $\mu\lr{\omega}$ (just like factorization into prime numbers is an NP-hard problem, but the validity of a given factorization can be checked in polynomial time). In particular, we conclude that the simple RTA prescription is not consistent with the QMC data and should be modified, for example, by considering the energy dependence of $\tau_{loc}$.

The codes used to produce the data that we refer to in this work are available at \href{https://github.com/j-ostmeyer/electron-phonon-hmc}{https://github.com/j-ostmeyer/electron-phonon-hmc} (QMC code), \href{https://github.com/buividovich/SSHExDiag}{https://github.com/buividovich/SSHExDiag} (exact diagonalization code) and \href{https://github.com/buividovich/TLPP}{https://github.com/buividovich/TLPP} (static approximation to $\mu\lr{\omega}$).

The authors are grateful to Simone Fratini and David Manolopolous for useful discussions. The work of J.~Ostmeyer was funded in part by the Deutsche Forschungsgemeinschaft (DFG, German Research Foundation) as part of the CRC 1639 NuMeriQS -- project no.\ 511713970. The work of P.~Buividovich was funded in part by the STFC Consolidated Grant ST/X000699/1.


\end{document}